\documentclass[twocolumn]{article}
\usepackage{algorithm}
\usepackage{algpseudocode}
\usepackage{bm}
\usepackage{graphicx}
\usepackage{subfigure}
\usepackage{tablefootnote}
\usepackage{url}

\begin{document}

\title{Length Matters: Clustering System Log Messages\\
  using Length of Words}
\author{Keiichi Shima\footnote{IIJ Innovation Institute, Inc.}}
\date{\empty}

\maketitle

\begin{abstract}
  The analysis techniques of system log messages (syslog messages)
  have a long history from when the syslog mechanism was invented.
  Typically, the analysis consists of two parts, one is a message
  template generation, and the other is finding something interesting
  using the messages classified by the inferred templates.  It is
  important to generate better templates to achieve better, precise,
  or convincible analysis results.  In this paper, we propose a
  classification methodology using the length of words of each
  message.  Our method is suitable for online template generation
  because it does not require two-pass analysis to generate template
  messages, that is an important factor considering increasing amount
  of log messages produced by a large number of system components such
  as cloud infrastructure.
\end{abstract}

%% \begin{CCSXML}
%% <ccs2012>
%% <concept>
%% <concept_id>10002951.10003227.10003351.10003444</concept_id>
%% <concept_desc>Information systems~Clustering</concept_desc>
%% <concept_significance>500</concept_significance>
%% </concept>
%% <concept>
%% <concept_id>10002978.10003006</concept_id>
%% <concept_desc>Security and privacy~Systems security</concept_desc>
%% <concept_significance>300</concept_significance>
%% </concept>
%% <concept>
%% <concept_id>10002978.10003014</concept_id>
%% <concept_desc>Security and privacy~Network security</concept_desc>
%% <concept_significance>300</concept_significance>
%% </concept>
%% </ccs2012>
%% \end{CCSXML}

%% \ccsdesc[500]{Information systems~Clustering}
%% \ccsdesc[300]{Security and privacy~Systems security}
%% \ccsdesc[300]{Security and privacy~Network security}

\section{Introduction}

The syslog mechanism and its protocols\cite{rfc3164,rfc5424} are
widely deployed in various kinds of systems to collect system status
messages from an informational level to a critical level in a
standardized way.  Since the original syslog protocol did not define
any message body structure, the log messages are typically in free
form text messages.  The newer syslog protocol
specification\cite{rfc5424} tried to organize semantic structures in
the body part of a message, however, not many programs respect the
specification so far.  Moreover, the decision on whether to use the newer
format or not depends on vendors, software, or even individual
programmers sometimes, we anyway have to handle both old and new
message formats.

There are various approaches to infer log message templates.
SLCT\cite{vaarandi2003-clustering} is one of the basic approach to
infer message templates without any prerequisite knowledge. SLCT is a
two-pass template inferring method.  In the first pass, it counts the
number of words that appear in the entire log messages and find
frequent words and its positions in a message.  The more frequently a
word appears, we can guess the word is likely a fixed keyword of the
template message.  For example, a message ``\texttt{interface eth0
  up}'' is covered by the template ``\texttt{\{(interface, 0), (up,
  2)\}}'' that means the keyword ``\texttt{interface}'' is at the
position 0, and ``\texttt{up}'' is at the position 2.

LogCluster\cite{vaarandi2015-logcluster} is similar to SLCT but
addressing shortcomings of SLCT.  SLCT creates templates as a set of
pairs of a word and its position.  Because of this, SLCT is sensitive
to the position of words.  LogCluster allows variable length of
parameters between fixed words.  For example, ``\texttt{interface eth0
  up}'' and ``\texttt{interface HQ Link up}'' are covered by one
template ``\texttt{interface *\{1,2\} up}'', where
``\texttt{*\{1,2\}}'' means 1 or 2 wildcard words.  Same as SLCT,
LogCluster requires a two-pass processing to detect the list of
frequent words.

Xu, el al. proposed a method using source code knowledge to infer
message templates in \cite{xu2009-mining-console-logs}. This is useful
when we know what kind of software are used in the target operation
system.  This approach requires preparation before classifying log
messages.  It also requires to update the inferred log template when
software used in the target system is added or updated.

Kimura, et al. introduced a character class based clustering method in
their work\cite{kimura2015-failure-detection}. They defined 5 classes
of words each consists of only numbers, numbers and letters, symbols
and letters, only letters, and only symbols respectively.  The latter
classes are considered more important than the former classes.  The
weight values for how much emphasize each class are pre-calculated
based on a PA-I supervised leaning algorithm\cite{crammer2006-pa-i}.
When comparing two messages, the ratio of the number of classes
included in each message is used.  The more the messages are similar,
the ratio will get closer to 1.

SHISO\cite{mizutani2013-mining-syslog-format} is another template
generation method focusing on online processing.  It calculates a
property of words as a vector by counting types of characters included
in a word such as capital alphabets, lower alphabets, numbers, marks,
and so on.  SHISO computes a Euclidean distance between words of two
messages being compared and generates similarity index of two
messages. If the index is smaller than the pre-defined threshold,
SHISO infers the two are similar and makes a cluster.

\section{Analysis of Messages in the Wild}

\subsection{Properties of Word Length in Messages}
\label{sec:properties-word-length}

Since syslog messages are printed by programs, each message has a
pre-formatted style.  \figurename\ref{fig:message-examples}
shows some examples of system log messages.

\begin{figure*}[!t]
  \centering
  \small
\begin{verbatim}
Oct  1 00:12:51 backup sshd[6854]: Invalid user vyatta from 41.190.192.158
Oct  1 00:12:51 backup sshd[6854]: input_userauth_request: Invalid user vyatta [preauth]
Oct  1 01:02:55 backup CRON[7069]: pam_unix(cron:session): session closed for user root
\end{verbatim}
  \caption{Examples of syslog messages.}
  \label{fig:message-examples}
\end{figure*}

As many previous works explained, a message comprises of two kinds of
components, one is a fixed component and the other is a variable
component.  In the first line in
\figurename\ref{fig:message-examples}, assuming that we know the head
of the message contains date information and host information,
\texttt{sshd}, \texttt{6845}, \texttt{vyatta}, and
\texttt{41.190.192.158} are variable components, while
\texttt{Invalid}, \texttt{user}, and \texttt{from} are fixed
components.  Many existing methods try to classify these components
based on some pre-defined knowledge, such as frequency of appearance,
ratio of character type, and so on.  Our simple question was that do
we really need to consider the property of each word.

\begin{figure*}[!t]
  \centering \small
\begin{verbatim}
Dec  1 00:05:01 vm1.example.com postfix/cleanup[2767]: 7EF561405E3:
message-id=<20151130150501.7EF561405E3@vm1.example.com>
Dec  1 00:10:01 vm1.example.com postfix/cleanup[3247]: 898FD1405E3:
message-id=<20151130151001.898FD1405E3@vm1.example.com>

Dec  1 00:27:27 backup sshd[15406]: Invalid user admin from 222.186.30.174
Dec  1 04:29:58 backup sshd[16287]: Invalid user a from 218.38.12.218
\end{verbatim}
  \caption{Examples of groups of syslog messages.}
  \label{fig:similar-message-examples}
\end{figure*}

\figurename\ref{fig:similar-message-examples} shows different examples
of similar messages that should be clustered into the same
group\footnote{Note that some messages are folded in the middle of the
  message due to the limitation of page width.}.  If we read these
messages, we can easily create a cluster of
``\texttt{postfix/cleanup[*]: *: message-id=*}'' for the first group,
and ``\texttt{sshd[*]: Invalid user * from *}'', where ``\texttt{*}''
means a variable component because we have knowledge of what is a
process identifier, or what is an IP address to infer which parts are
fixed and which are not.  But even though we do not use such
knowledge, here is another factor we can read from these examples,
that is the length of each word.  It is obvious that all the fixed
components have the same word length in the message.  Variable words
have different word length, but tend to have similar length because
they share the same context, such as a message identifier, an IP
address, a process identifier, a host/user name, and so on.

\begin{figure*}[!t]
  \centering
  \subfigure[\texttt{sshd * : Received disconnect from * 11: * * *}
    (79182 samples)]{
    \includegraphics[width=0.3\textwidth]{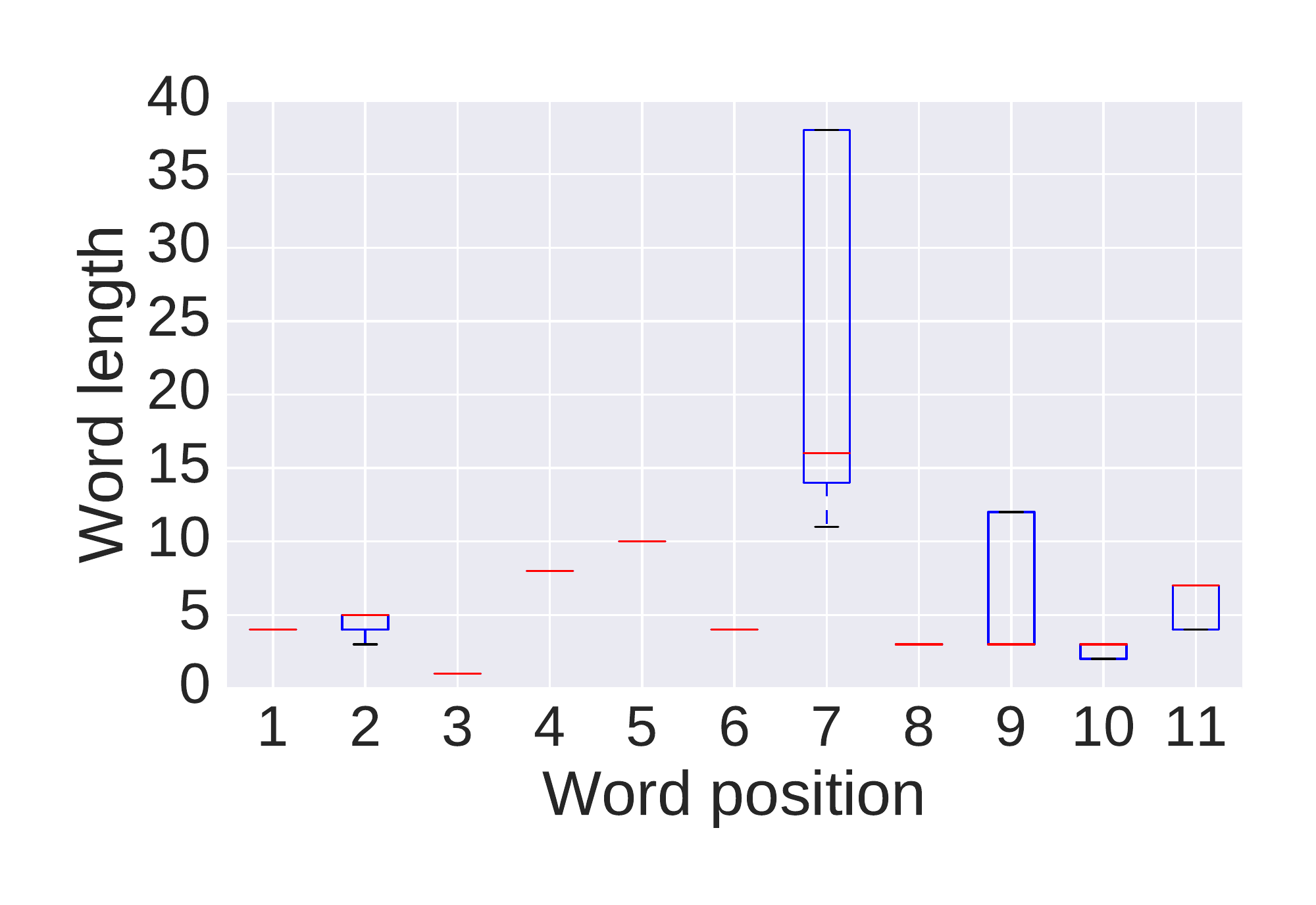}
    \label{subfig:sshd-1}
  }
  \hspace{0.5em}
  \subfigure[\texttt{sshd * : reverse mapping checking getaddrinfo
      for * * failed - POSSIBLE BREAK-IN ATTEMPT!} (11926 samples)]{
    \includegraphics[width=0.3\textwidth]{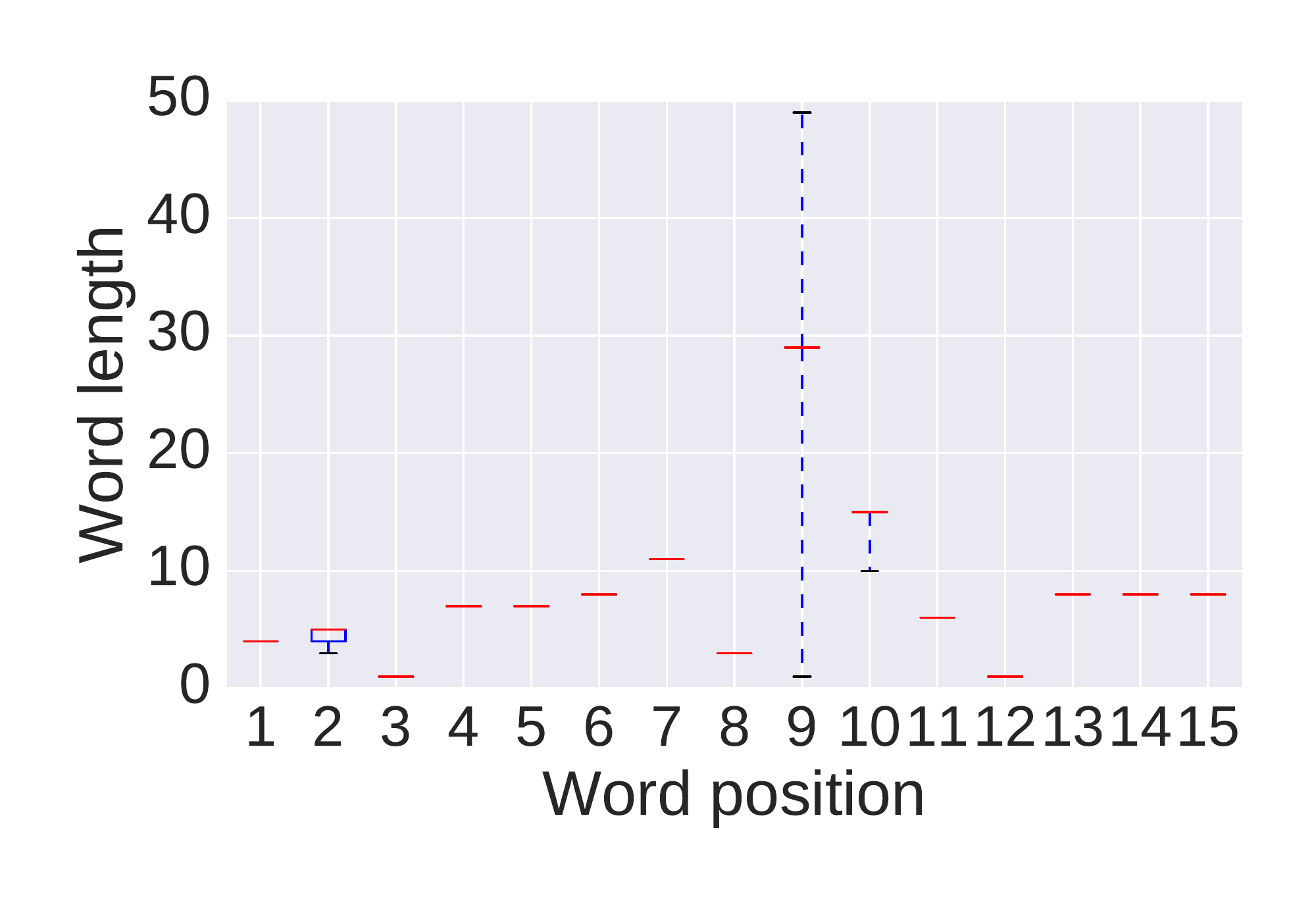}
    \label{subfig:sshd-2}
  }
  \hspace{0.5em}
  \subfigure[\texttt{named * : DNS format error from * resolving *
      invalid response} (878 samples)]{
    \includegraphics[width=0.3\textwidth]{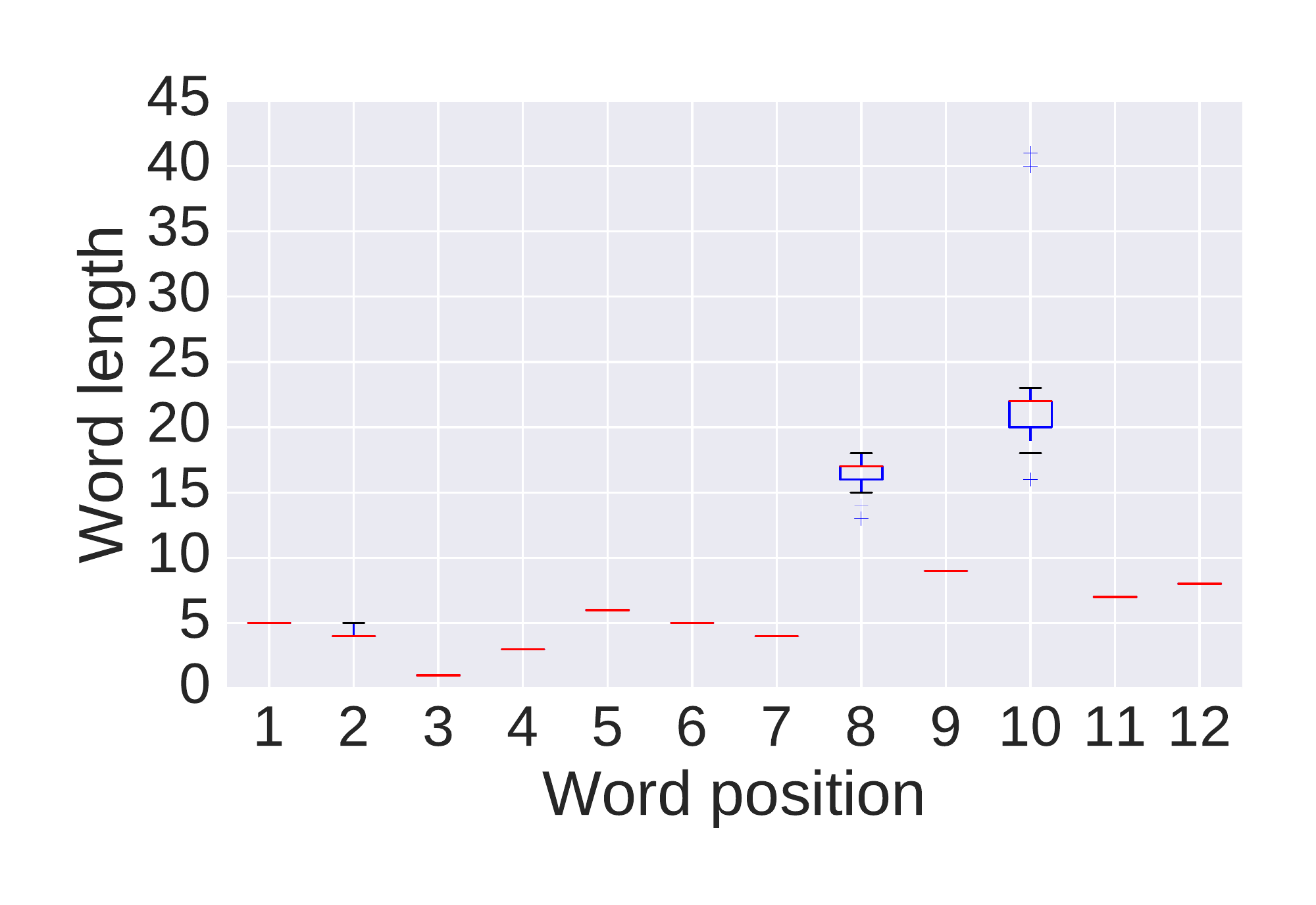}
  }
  \caption{Examples of word length distribution patterns of some
    syslog messages extracted from the messages on 1st October 2015 of
    the dataset \#2 in Table~\ref{tab:syslog-datasets}.}
  \label{fig:length-distribution-examples}
\end{figure*}

\figurename\ref{fig:length-distribution-examples} shows examples of
distribution of word length of some messages groups.  As we can read
from the figures, each syslog message has a unique pattern of
distribution of length of each words.  The length of the first
position is usually fixed because a process name is printed here
normally.  The second position is process identifiers and it is
usually 3 to 5 digits.  The 7th position of
\figurename\ref{subfig:sshd-1} is a placeholder for IP addresses and
contains either IPv4 or IPv6 address.  Because IPv6 address can be
printed shorter by eliminating zero fields, the position has wider
range of length.  The 9th position of \figurename\ref{subfig:sshd-2}
is a placeholder for host names.  In our data, the median of the length
of the position was 29 and almost fixed however, we saw some very short
and long host names in the log.

\begin{figure}[!t]
  \centering
  \includegraphics[width=0.4\textwidth]{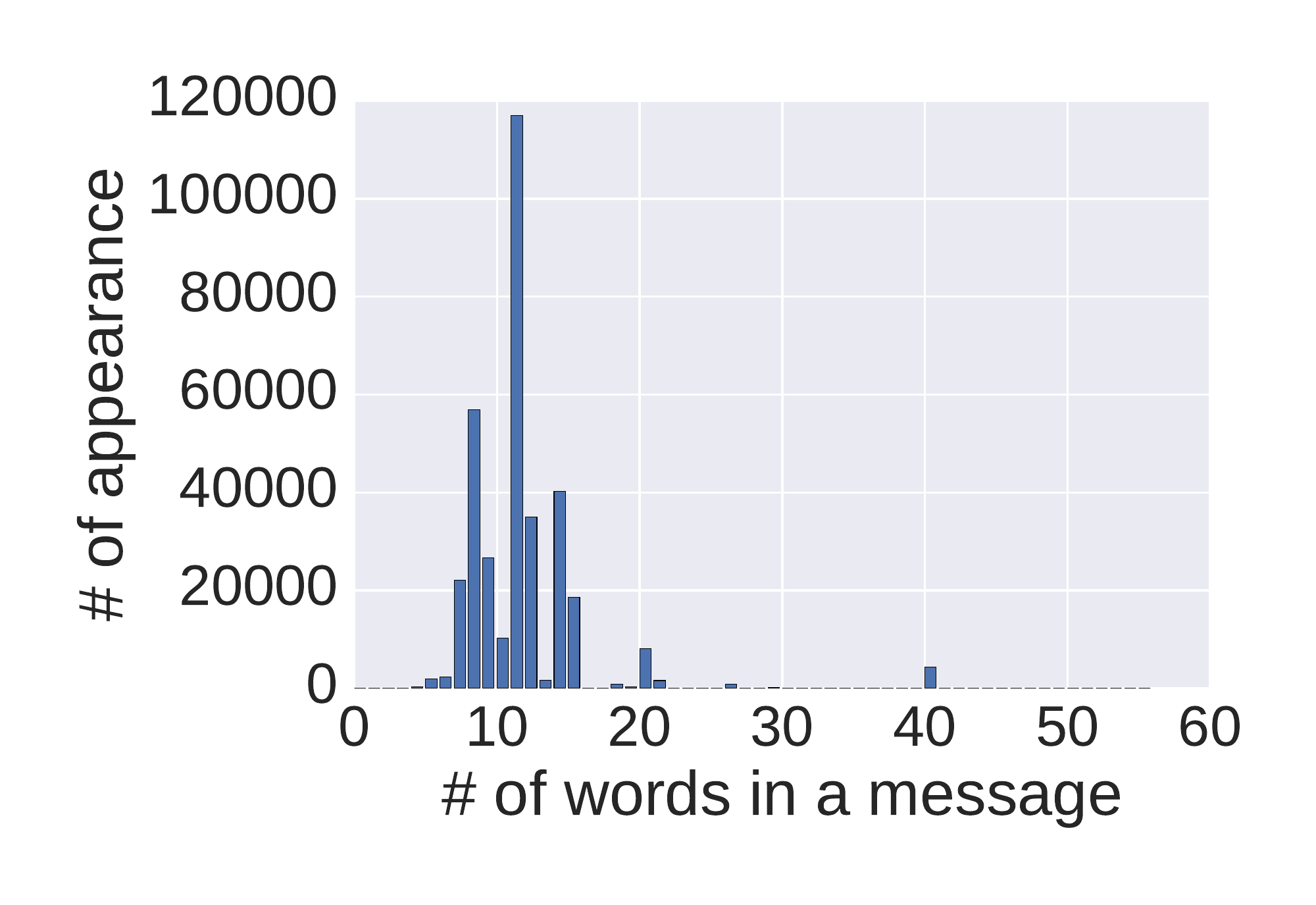}
  \caption{The distribution of the number of words of syslog messages
    on 1st October 2015 of the dataset \#2 in
    Table~\ref{tab:syslog-datasets}.}
  \label{fig:nwords-distribution}
\end{figure}

We analyzed how much the set of length of words are correlated each
other using the syslog message dataset \#2 shown in
Table~\ref{tab:syslog-datasets}.  The dataset is a collection of
messages gathered from hypervisors operated by the WIDE
project\footnote{\url{http://www.wide.ad.jp/}}.  The distribution
of the number of words of each message (excluding date and host name
information) is shown in \figurename\ref{fig:nwords-distribution}.

\begin{figure}[!t]
  \centering
  \includegraphics[width=0.4\textwidth]{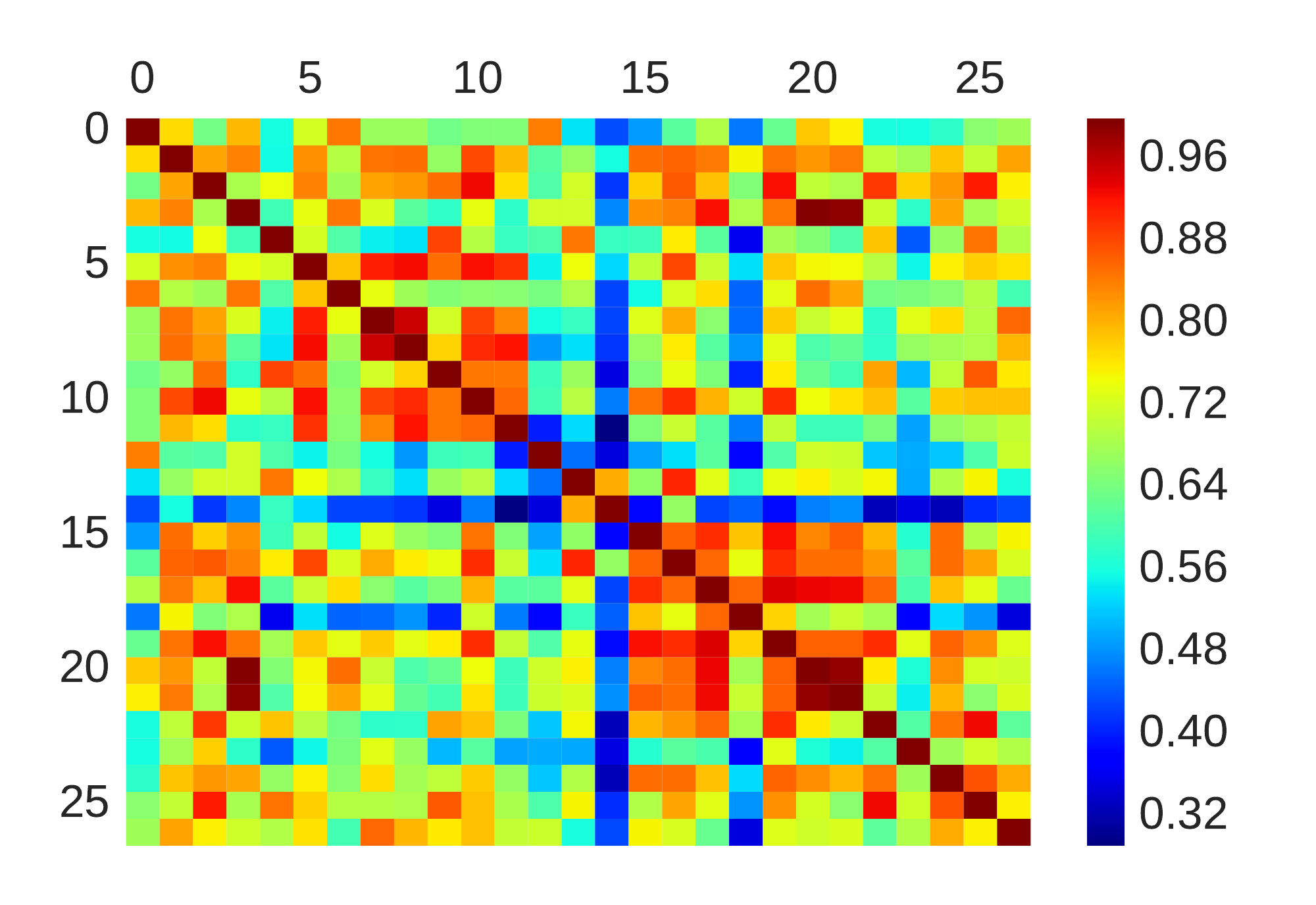}
  \caption{The similarity between two message templates shown in
    Table~\ref{tab:log-group-for-correlation} using the cosine
    similarity over the vectors calculated by converting messages to
    vectors of the length of each word.}
  \label{fig:11-words-correlation}
\end{figure}

\begin{table*}[!t]
  \footnotesize
  \renewcommand{\arraystretch}{1.3}
  \caption{Message Templates of Syslog Messages with 11 Words
    Extracted Half-Manually from the Messages on 1st October 2015 of
    the Dataset \#2 in Table~\ref{tab:syslog-datasets}.}
  \label{tab:log-group-for-correlation}
  \centering
  \begin{tabular}{r|l|r}
    \hline
    \textbf{\#} & \textbf{Template} & \textbf{\# of msgs}\\
    \hline\hline
    0 & \verb|40grub2: debug: parsing: if $linux_gfx_mode ! text ; then load_video; fi| & 5\\
    \hline
    1 & \verb|40grub2: debug: parsing: menuentry Memory test memtest86+, serial console 115200 {| & 5 \\
    \hline
    2 & \verb|40grub2: appears to be an automatic reference taken from another menu.lst| & 14 \\
    \hline
    3 & \verb|CRON * : pam_unix cron:session : session closed for user *| & 8562 \\
    \hline
    4 & \verb|CRON * : root CMD cd / && run-parts --report /etc/cron.hourly| & 768 \\
    \hline
    5 & \verb|kernel: * Buffer I/O error on device loop0p1, logical block *| & 10 \\
    \hline
    6 & \verb|kernel: * INFO: task * blocked for more than 120 seconds.| & 12 \\
    \hline
    7 & \verb|kernel: * init: * main process * terminated with status *| & 18 \\
    \hline
    8 & \verb|kernel: 173315.040098 EXT4-fs vda1 : error count since last fsck: 3| & 1 \\
    \hline
    9 & \verb|kernel: * EXT4-fs vda1 : * error at time * *| & 2 \\
    \hline
    10 & \verb|kernel: * init: Failed to obtain startpar-bridge instance: Unknown parameter: INSTANCE| & 2 \\
    \hline
    11 & \verb|kernel: * systemd-logind 2127 : New session * of user * | & 1080 \\
    \hline
    12 & \verb|named * : client * view world: query cache * denied| & 3120 \\
    \hline
    13 & \verb|named * : error unexpected RCODE * resolving * : *| & 2224 \\
    \hline
    14 & \verb|ntpd * : Listen normally on * * * UDP 123| & 29 \\
    \hline
    15 & \verb|sshd * : Disconnecting: Too many authentication failures for * preauth| & 409 \\
    \hline
    16 & \verb|sshd * : User * not allowed because account is locked| & 1003 \\
    \hline
    17 & \verb|sshd * : Received disconnect from * 11: Bye Bye preauth| & 48299 \\
    \hline
    18 & \verb|sshd * : Received disconnect from * 11: disconnected by user| & 30883 \\
    \hline
    19 & \verb|sshd * : fatal: Write failed: Connection reset by peer preauth| & 21 \\
    \hline
    20 & \verb|sshd * : pam_unix sshd:session : session closed for user *| & 20525 \\
    \hline
    21 & \verb|su * : pam_unix su:session : session closed for user *| & 9 \\
    \hline
    22 & \verb|postfix/flush * : fatal: config variable inet_interfaces: host not found: *| & 5 \\
    \hline
    23 & \verb|postfix/master * : warning: process /usr/lib/postfix/flush pid * exit status 1| & 5 \\
    \hline
    24 & \verb|postfix/smtpd * : SSL_accept error from unknown * : lost connection| & 1 \\
    \hline
    25 & \verb|postfix/smtpd * : too many errors after DATA from unknown *| & 1 \\
    \hline
    26 & \verb|rpcbind: connect from * to dump : request from unauthorized host| & 14 \\
    \hline
  \end{tabular}
\end{table*}
The most popular message group was those whose number of words was 11.
We then made 27 message templates by half-manual way to split the
messages as shown in
Table~\ref{tab:log-group-for-correlation}. \figurename\ref{fig:11-words-correlation}
shows the similarity matrix between two templates of
Table~\ref{tab:log-group-for-correlation}.  We found that it would be
possible to distinguish two templates by measuring distance between
them in most cases.

We also found that in some cases, the distance of
two template messages is quite small even though the messages are
completely different.  We analyze such cases in
Section~\ref{sec:properties-word-order}.

\subsection{Properties of Positions of Words in Messages}
\label{sec:properties-word-order}

The one of the primitive clustering methods of syslog messages is
creating clusters based on the number of words in the messages.  Since
the messages are printed based on pre-defined styles, the resulting
messages will have the same number of words usually.  Because this
clustering method is too primitive, we need to create sub-clusters
using other information.  In the previous subsection, we have
mentioned that each syslog message cluster tends to have a similar
series of word length values.  This observation helps to make
sub-clusters in a cluster that has the same number of words in
messages, however, there is some cases we need a different index other
than word length values.

\begin{figure*}[!t]
  \centering
  \subfigure[\texttt{sshd * : refused connect from * *}
    (762 samples)]{
    \includegraphics[width=0.3\textwidth]{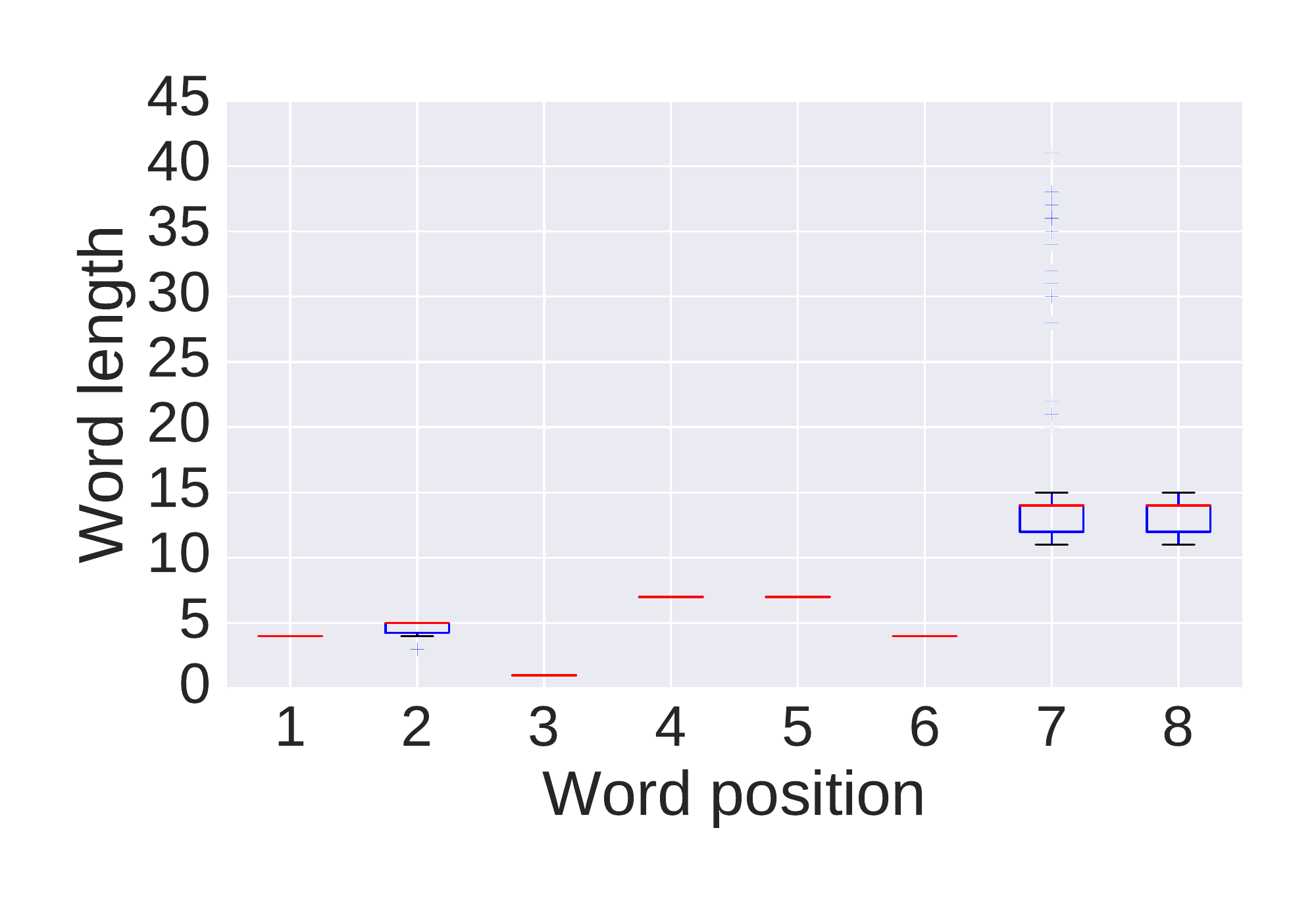}
  }
  \hspace{0.5em}
  \subfigure[\texttt{sshd * : Connection closed by * preauth}
    (16192 samples)]{
    \includegraphics[width=0.3\textwidth]{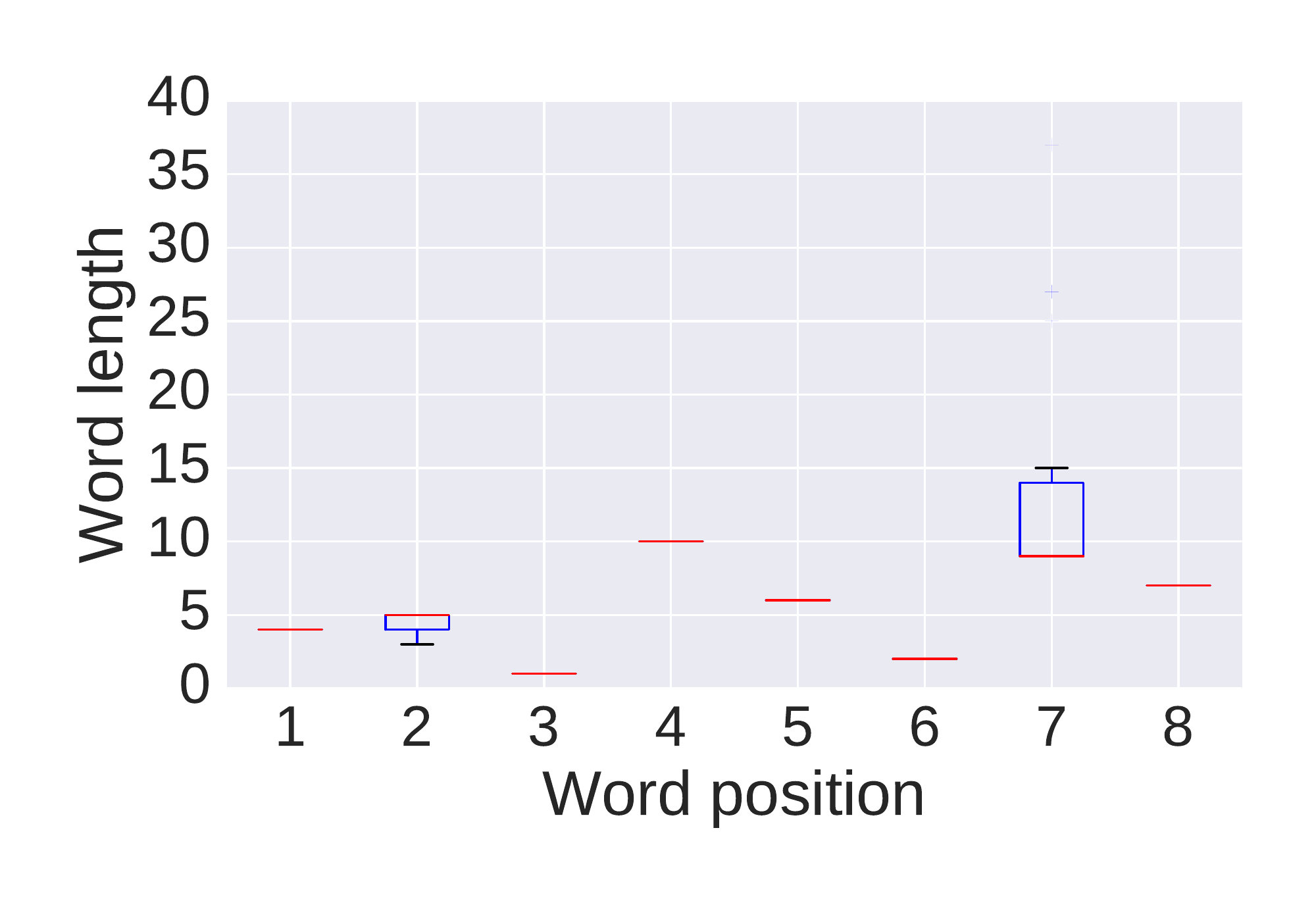}
  }
  \caption{Examples of distribution patterns of word length of two
    completely different syslog messages where the distance between
    them becomes close.}
  \label{fig:two-similar-vectors}
\end{figure*}

\figurename\ref{fig:two-similar-vectors} shows the word length
distribution patterns of two different syslog messages.  In the later
Section~\ref{sec:lenma}, we will discuss how to compare similarity
between existing clusters and an incoming syslog message in detail,
but in short, we use \textit{Cosine Similarity} as a base idea for
comparison.  However, the two messages shown in
\figurename\ref{fig:two-similar-vectors} are quite similar in the
sense of cosine similarity.

It is obvious for us to conclude that these two messages are
different.  The two messages in
\figurename\ref{fig:two-similar-vectors} only share the first word
(and the third word ``\texttt{:}'') which is the name of the process
that wrote the messages.  The rest of the fixed components of the
messages are completely different.  So we also focus on the word
positions of messages.  If the words of two messages being compared
doesn't have shared words in the same position, then we can think they
should be clustered to different groups.

\section{LenMa: Length Matters Clustering}
\label{sec:lenma}

Based on the observation in Section~\ref{sec:properties-word-length},
we focus on the length of each words of the message as a similarity
parameters of the message.

When clustering messages online, we need to compare the latest
incoming message and existing clusters to decide which is the best
cluster that should include the incoming message or create a new
cluster if none of the existing clusters suites the message.

For the first line of the first group in
\figurename\ref{fig:similar-message-examples}, the length of words can
be represented as a vector of
\begin{eqnarray}
  [len(\texttt{postfix/cleanup}), \nonumber \\
    len(\texttt{2767}), \nonumber \\
    len(\texttt{7EF561405E3}), \nonumber \\
    len(\texttt{message-id}), \nonumber \\
    len(\texttt{<201511}...(snip)...\texttt{example.com>})] \nonumber \\
  = [15, 4, 11, 10, 44] \nonumber
\end{eqnarray}
which is the same vector calculated from the second line of the first
group.  For the second group, the vectors are $[4, 5, 7, 4, 5, 4, 13]$
and $[4, 5, 7, 4, 1, 4, 13]$.  If the vectors of two messages are
similar, then we can guess that these messages are in a same cluster.

The similarity $S_{c}$ is calculated from the word length vectors of
the existing cluster and incoming new message as shown in
(\ref{eq:similarity}) using the cosine similarity.

\begin{eqnarray}
  \bm{V}_{c} & =  & [v_{c,0}, v_{c,1}, ... v_{c,n}] \nonumber \\
  \bm{V} & = & [v_{0}, v_{1}, ... v_{n}] \nonumber \\
  S_{c} & = & CosineSimilarity(\bm{V}_{c}, \bm{V}) \nonumber \\
  & = & \frac{\bm{V}_{c}\cdot \bm{V}}{|\bm{V}_{c}|~|\bm{V}|} \nonumber \\
  & = & \frac{\sum_{i=0}^{n} v_{c,i}v_{i}}{\sqrt{\sum_{i=0}^{n}v_{c,i}^{2}}\sqrt{\sum_{i=0}^{n}v_{i}^{2}}}
  \label{eq:similarity}
\end{eqnarray}
where $\bm{V}_{c}$ and $\bm{V}$ are the word length vectors of the
cluster $c$ and incoming message respectively.  $v_{c,i}$ and $v_{i}$ are
the length of $i$th word of the cluster $c$ and incoming message.

Cluster vectors are updated whenever a new message is integrated to
existing clusters.  A new word length vector is calculated as shown in
Algorithm~\ref{alg:update-word-length-vector}.  If the length of the
$i$th word of the cluster and a new message is same, the value is kept
unchanged, otherwise, the length value is updated with the new length
value of the $i$th word of the new message.

Similarly, A new word vector which keeps a template string of the
cluster is updated as shown in Algorithm~\ref{alg:update-word-vector}.
$\bm{W}_{c}$ and $\bm{W}$ are ordered set of words of a cluster and
incoming message.  For example, the top message of
\figurename\ref{fig:message-examples} can be represented as
               [\texttt{sshd}, \texttt{6854}, \texttt{Invalid},
                 \texttt{user}, \texttt{vyatta}, \texttt{from},
                 \texttt{41.190.192.158}]. For the case of a cluster,
               some of the words are replaced with a wildcard mark as
               variables such as [\texttt{sshd}, \texttt{*},
                 \texttt{Invalid}, \texttt{user}, \texttt{*},
                 \texttt{from}, \texttt{*}].  When updating the word
               vector $\bm{W}_{c}$, if the $i$th word of the cluster
               $c$ is different from the $i$th word of the input word
               vector $\bm{W}$, the word is replaced by a wildcard
               mark, otherwise the value is kept unchanged.

\begin{algorithm}
  \caption{Update a word length vector}
  \label{alg:update-word-length-vector}
  \begin{algorithmic}
\Procedure{UpdateWordLengthVector}{$\bm{V}_{c}, \bm{V}$}
  \ForAll {$v_{c,i} \in \bm{V}_{c}, v_{i} \in \bm{V} (i \leftarrow 1\cdots|\bm{V}_{c}|)$}
    \If {$v_{c,i} \neq v_{i}$}
      \State $v_{c,i} \leftarrow v_{i}$
    \EndIf
  \EndFor
\EndProcedure
  \end{algorithmic}
\end{algorithm}

\begin{algorithm}
  \caption{Update a word vector}
  \label{alg:update-word-vector}
  \begin{algorithmic}
\Procedure{UpdateWordVector}{$\bm{W}_{c}, \bm{W}$}
  \ForAll {$w_{c,i} \in \bm{W}_{c}, w_{i} \in \bm{W} (i \leftarrow 1\cdots|\bm{W}_{c}|)$}
    \If {$w_{c,i} \neq w_{i}$}
      \State $w_{c,i} \leftarrow \textrm{*}$
    \EndIf
  \EndFor
\EndProcedure
  \end{algorithmic}
\end{algorithm}

When we receive a new log message, the following procedure is
executed.
\begin{enumerate}
\item Create a word length vector and word vector of the new message.
\item Calculate a similarity score between the new message and each
  cluster which has the same number of words.
\item If none of the cluster has a similarity value larger than the
  threshold $T_{c}$, a new cluster with the new message is created and
  returned.
\item Update the most similar cluster with the new message using the
  algorithms defined in Algorithm~\ref{alg:update-word-length-vector}
  and Algorithm~\ref{alg:update-word-vector}.
\item Return the most similar cluster.
\end{enumerate}

Algorithm~\ref{alg:find-cluster} shows the above procedure.  As we
discussed in Section~\ref{sec:properties-word-order}, we use the
cosine similarity score to compare the new message and existing
clusters, however it is not enough to determine a proper cluster in
some cases.  We need to judge if the fixed parts of the cluster
template are similar enough to the fixed parts of the new message.  To
achieve this goal, we introduced a positional similarity index $S_{p}$
based on the number of shared words at the same positions.  The
$S_{p}$ is calculated as shown in (\ref{eq:positional-similarity}).
\begin{eqnarray}
  S_{p} &=& |\{w_{c,i} = w_{i} (w_{c,i} \in \bm{W}_{c}, w_{i} \in \bm{W})\}|
  \label{eq:positional-similarity}
\end{eqnarray}
where $\bm{W}_{c}$ and $\bm{W}$ are the word vectors of the cluster
and incoming message, $i$ is the position of the word in a template or
a message.

When comparing a cluster template and incoming message, we consider how
many shared words are there in the same positions.  If the number is
smaller than the pre-defined threshold $T_p$, the message is
considered as out of the cluster.

\begin{algorithm}
  \caption{Find or Create a cluster}
  \label{alg:find-cluster}
  \begin{algorithmic}
\State $\bm{C} \leftarrow \emptyset$
\Procedure{FindOrCreateCluster}{$message$}
  \State $\bm{V} \leftarrow$ \Call{CreateWordLengthVector}{$message$}
  \State $\bm{W} \leftarrow$ \Call{CreateWordVector}{$message$}
  \State $\bm{Cand} \leftarrow \emptyset$
  \ForAll {$\left[\bm{V}_{c}, \bm{W}_{c}\right]$ in $\bm{C}$}
    \If {\Call{Similarity}{$\bm{V}_{c}, \bm{V}, \bm{W}_{c}, \bm{W}$} $> T_{c}$}
      \State {$\bm{Cand} \leftarrow \bm{Cand} \cap \left[\bm{V},\bm{W}\right]$}
    \EndIf
  \EndFor
  \If {$\bm{Cand} = \emptyset$}
    \State {$\bm{C} \leftarrow \bm{C} \cap \left[\bm{V},\bm{W}\right]$}
    \State \Return $\left[\bm{V}, \bm{W}\right]$
  \EndIf
  \State $\left[\bm{V}_{c},\bm{W}_{c}\right] \leftarrow$ \Call{HighestSimilarity}{$\bm{Cand}$}
  \State \Call{UpdateWordLengthVector}{$\bm{V}_{c}, \bm{V}$}
  \State \Call{UpdateWordVector}{$\bm{W}_{c}, \bm{W}$}
  \State \Return $\left[\bm{V}_{c}, \bm{W}_{c}\right]$
\EndProcedure
  \end{algorithmic}
\end{algorithm}

\begin{algorithm}
  \caption{Calculate similarity}
  \label{alg:calculate-similarity}
  \begin{algorithmic}
\Procedure{Similarity}{$\bm{V}_{c}, \bm{V}, \bm{W}_{c}, \bm{W}$}
  \If {$|\bm{V}_{c}| \neq |\bm{V}|$}
    \State \Return $0$
  \EndIf
  \If {$\bm{W}_{c}$ matches $\bm{W}$}
    \State \Return $1$
  \EndIf
  \State $S_{c} \leftarrow$ \Call{CosineSimilarity}{$\bm{V}_{c}, \bm{V}$}
  \State $S_{p} \leftarrow |\{w_{c,i} = w_{i} (w_{c,i} \in \bm{W}_{c}, w_{i} \in \bm{W})\}|$
  \If {$S_{p} < T_{p}$}
    \State \Return $0$
  \EndIf
  \State \Return $S_{c}$
\EndProcedure
  \end{algorithmic}
\end{algorithm}

\section{Implementation}

\begin{table}[!t]
  \renewcommand{\arraystretch}{1.3}
  \caption{Syslog Datasets}
  \label{tab:syslog-datasets}
  \centering
  \begin{tabular}{c|p{5.5cm}}
    \hline
    \textbf{Dataset \#} & \textbf{Content of dataset} \\
    \hline\hline
    1 & Public Security Log Sharing Site\cite{chuvakin2009} \\
    \hline
    2 & Hypervisor cluster operated by the WIDE project\\
    \hline
    3 & Server cluster of our laboratory\\
    \hline
  \end{tabular}
\end{table}

We have implemented the proposed algorithm in Python and applied it
with three different syslog datasets shown in
Table~\ref{tab:syslog-datasets}.  The dataset \#1 is a public data
provided by Chuvakin\cite{chuvakin2009}.  The dataset \#2 is a set of
syslog messages collected at the hypervisor cluster operated by the
WIDE project.  The dataset \#3 is a set of syslog messages collected
from servers of our laboratory including several hypervisors and
service hosts such as web servers.

We introduced one heuristic approach assuming the beginning of a
message includes a date string, a host name, and a process name that
is syslog specific.  Although RFCs says that the standard syslog
message starts with the date information followed by a host name, a
process name, and other components, we understand not all the syslog
messages doesn't strictly follow the recommended message
format. From our experience however, the first 3 components (date,
host name, and process name) share the same context in most cases.

\begin{figure}[!t]
  \centering
  \includegraphics[width=0.4\textwidth]{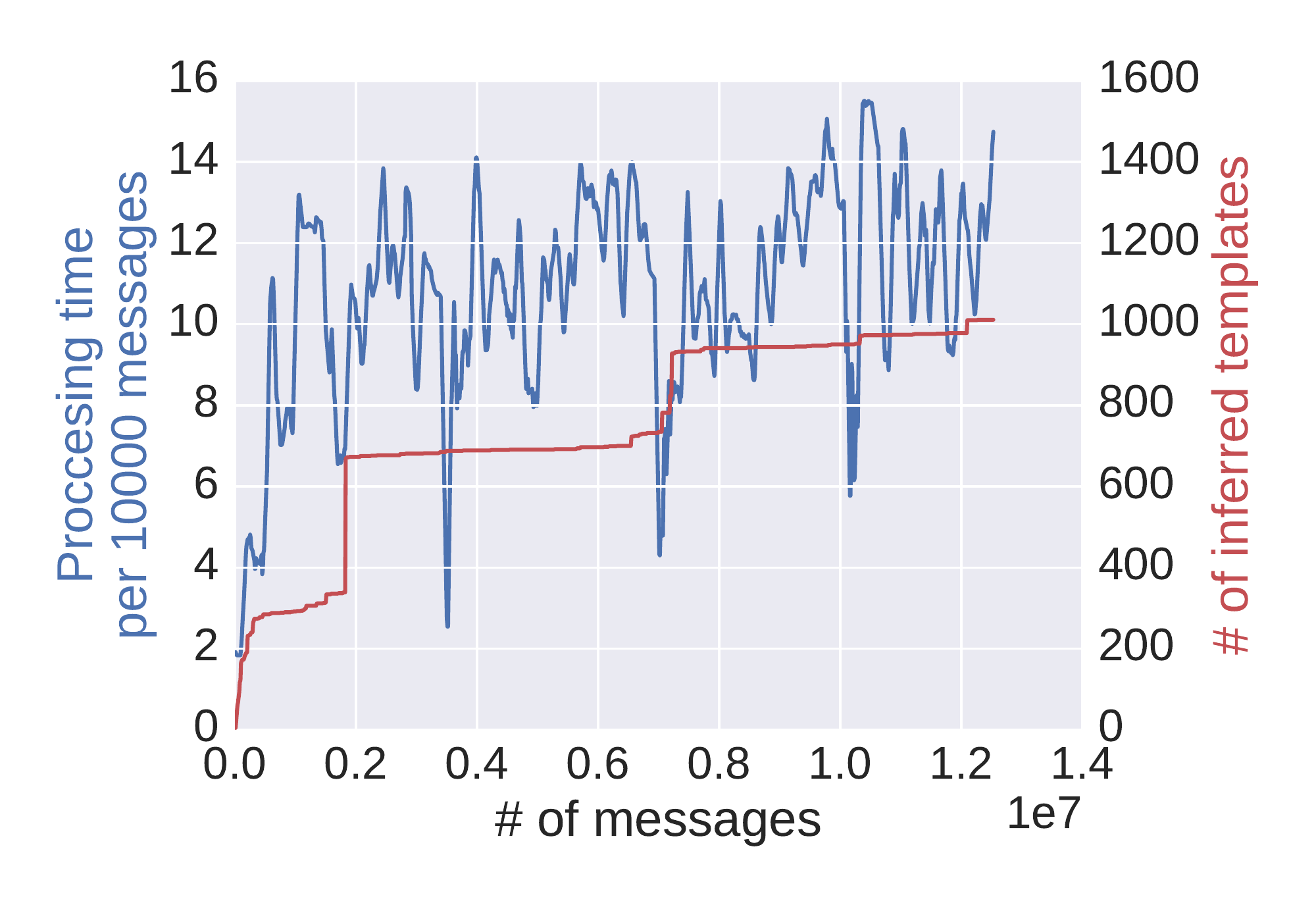}
  \caption{A processing time convergence graph measured with the messages
    in October 2015 of the dataset \#2.}
  \label{fig:processing-speed}
\end{figure}

The processing cost of inferring template formats linearly depends on
the number of inferred templates.  As shown in the algorithm, the
value of the threshold affects the final number of inferred templates.
The smaller threshold generates less number of templates that contain
more wildcard marks.  Once the number of templates becomes stable, the
algorithm can process messages within a certain fixed rate.
\figurename\ref{fig:processing-speed} shows the processing time of
messages in October 2015 of the dataset \#2.  The blue line indicates
the time required to process 10000 messages and the red line shows the
number of inferred templates.  The processing time increases as the
number of templates increases, however the number of the templates
becomes stable once we have found most of them and the processing time
also becomes stable.

We applied our algorithm and SHISO algorithm to cluster the datasets
shown in Table~\ref{tab:syslog-datasets}.  The result is shown in
Table~\ref{tab:clustering-results}.  We found a lot of kernel messages
that appear when a system boots up in the dataset \#2 and \#3.  In
\figurename\ref{fig:processing-speed}, such template messages are
found at 2 points, the first point is the point around 200 million
messages are processed, and the second point is where 700 million
message are processed.  We found a lot of kernel boot messages in the
inferred templates.  These messages appear only once while the system
is running but generates a lot of different patterns that increases
the total number of templates.  The values in parentheses are the
number of templates that are not related to kernel boot messages.
This indicates that the online template mining mechanisms work well
even for the infrequent messages, however cleansing of raw messages
may be required to avoid unwanted template generation.

\begin{table}
  \renewcommand{\arraystretch}{1.3}
  \caption{Clustering Results}
  \label{tab:clustering-results}
  \centering
  \begin{tabular}{l|r|r}
    \hline
    \textbf{Dataset \#} & \multicolumn{2}{c}{\textbf{\# of templates found}} \\
    & \textbf{SHISO}\tablefootnote{SHISO proposes a second level clustering that merges templates that contains similar set of words regardless the \# of words, however, in this paper we compare the result of the first level clustering result.} & \textbf{LenMa} \\
    \hline\hline
    1 (\texttt{log/secure*} only) & 29 & 26\\
    \hline
    2 & 1093 (446) & 1075 (404)\\
    \hline
    3 & 1113 (361) & 891 (302)\\
    \hline
  \end{tabular}
\end{table}

The prototype code is available at GitHub\footnote{
\url{https://github.com/keiichishima/templateminer}}.

\section{Using Clustered Syslog Messages for Analysis}

\begin{figure}[!t]
  \centering
  \includegraphics[width=\columnwidth]{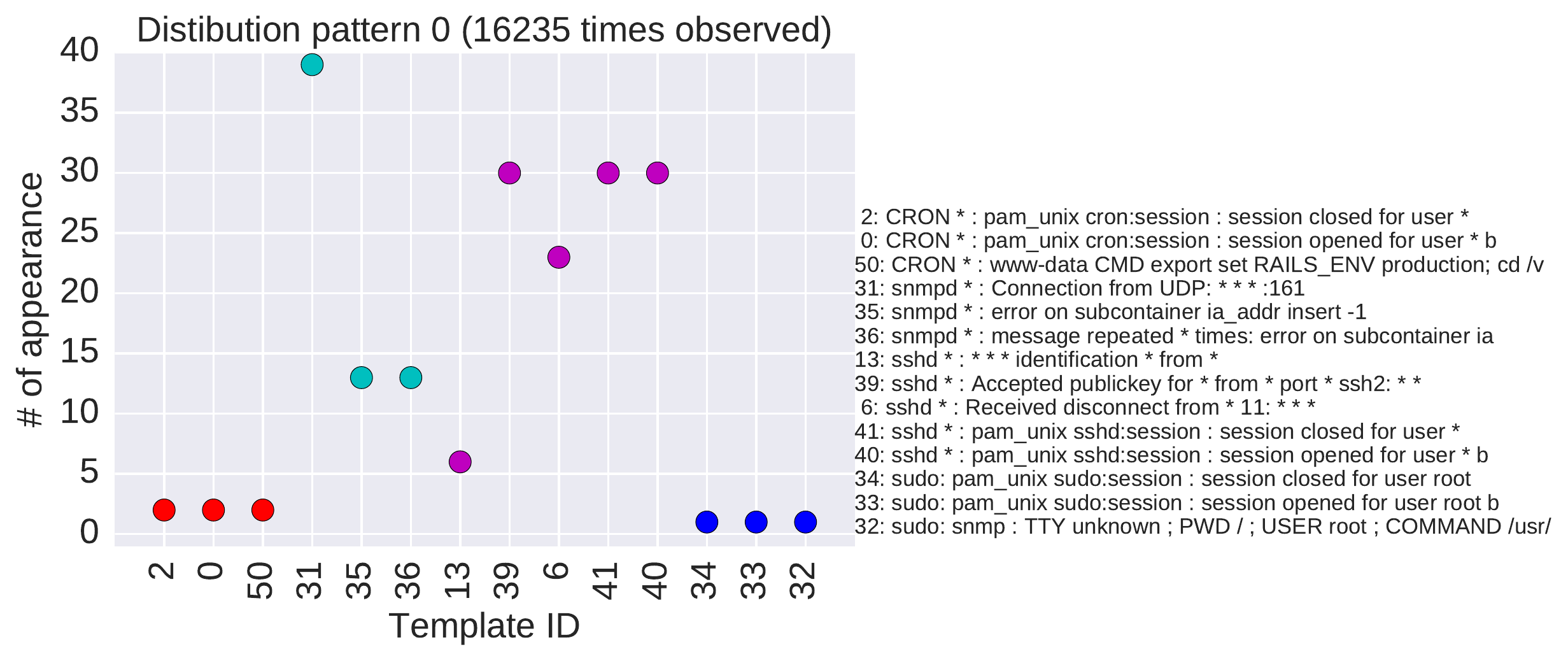}
  \includegraphics[width=\columnwidth]{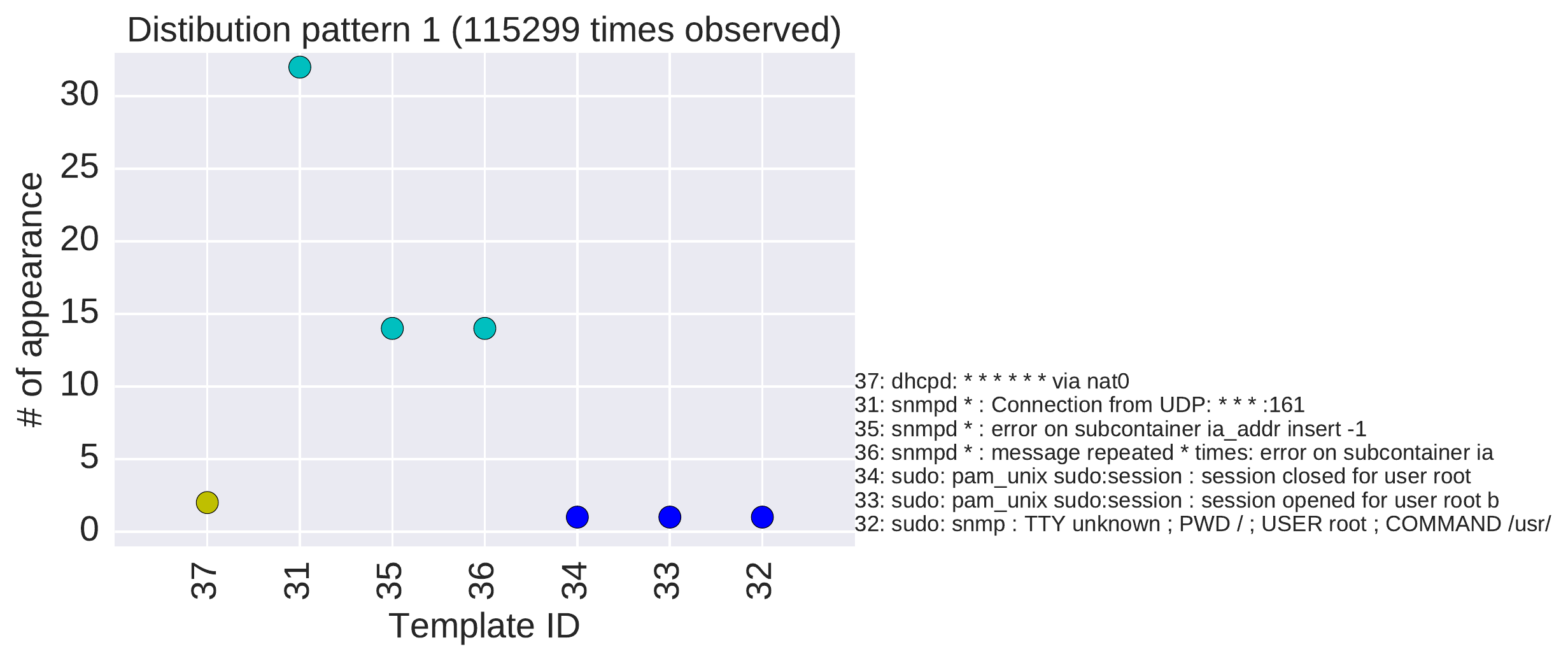}
  \caption{Frequently observed message group patterns where each color
    represents a process name group}
  \label{fig:frequent-patterns}
\end{figure}

We tried to find unique syslog message patterns using the messages
clustered by LenMa.  We clustered all the messages using the algorithm
and made message groups for every minute.  The dataset used for this
grouping is the dataset \#3 and the period is from October to December
2015.  Each group has its own distribution pattern of templates,
however, many of them are similar each other.  We counted the number
of appearance of templates of each group, clustered them using the
$\chi^2$ test, and finally achieved 25 message group clusters out of
132480 (= 60 minutes $\times$ 24 hours $\times$ 92 days) groups. The
frequently observed patterns are shown in
\figurename\ref{fig:frequent-patterns}, and the counts are 16235 and
115299 times respectively.  Since each group is one-minute-long,
almost 91 days out of 92 days match these patterns.

\begin{figure}[!t]
  \centering
  \subfigure[Unique ssh incoming activities observed]{
    \includegraphics[width=\columnwidth]{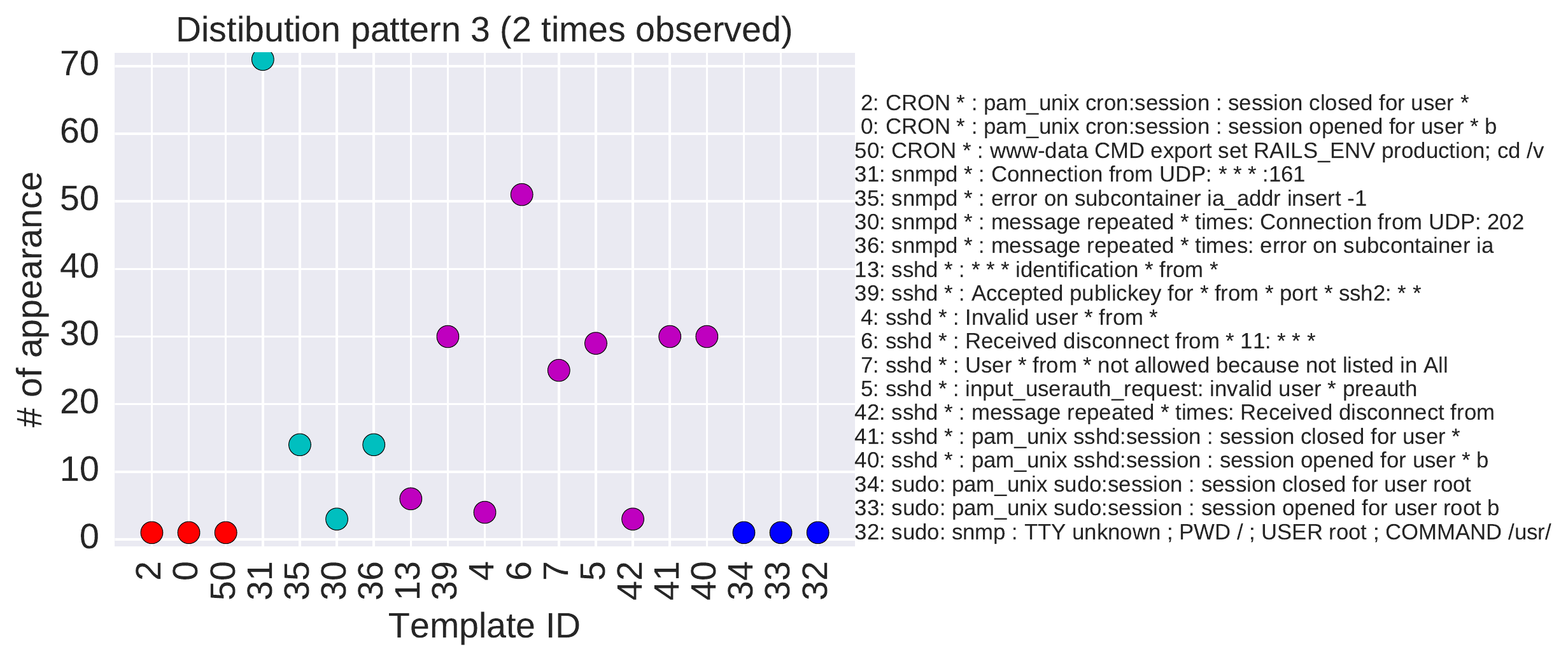}
  }
  \subfigure[Node rebooting observed]{
    \includegraphics[width=\columnwidth]{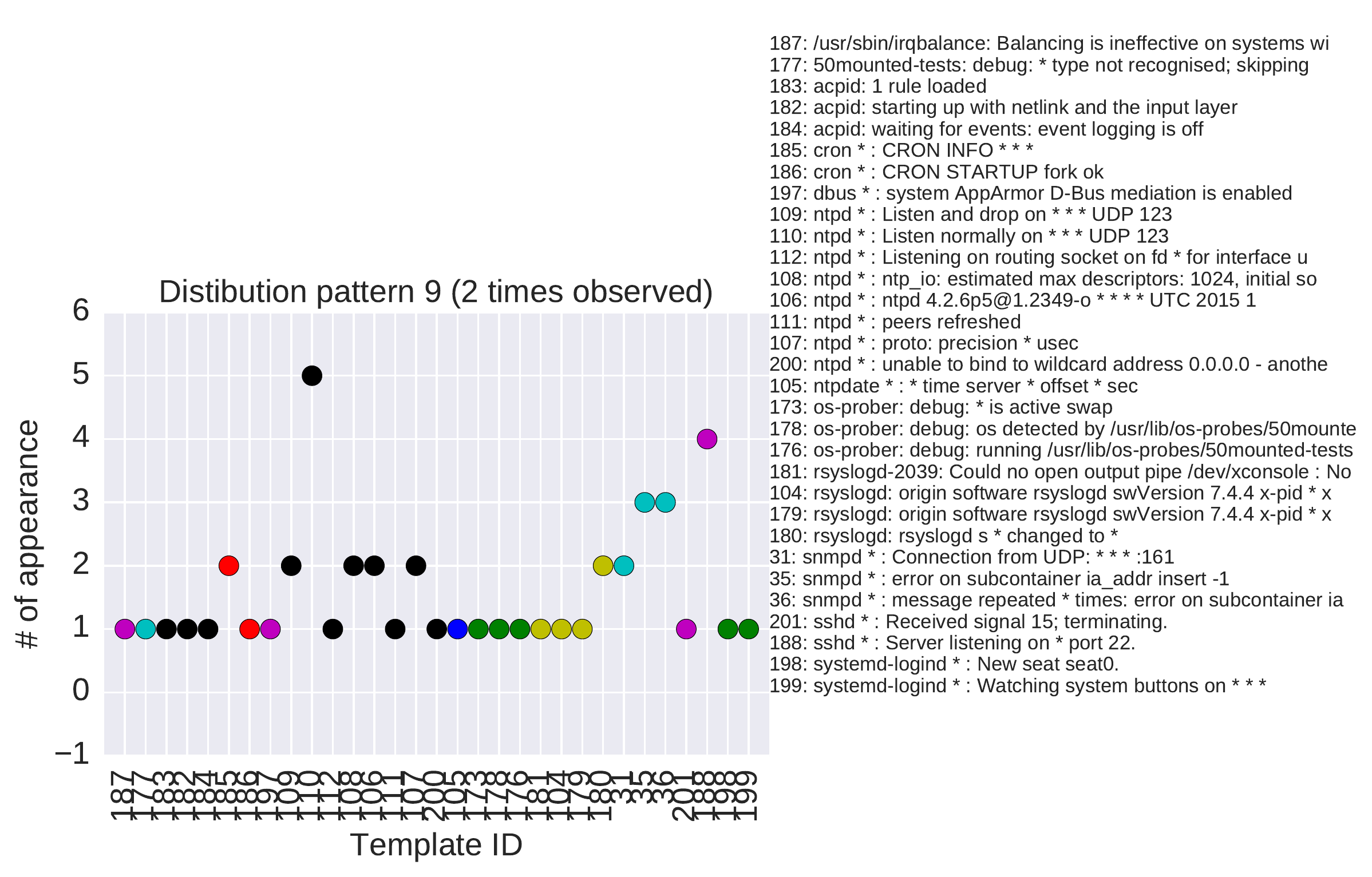}
  }
  \caption{Examples of unique message group patterns observed}
  \label{fig:unique-patterns}
\end{figure}

There were some unique patterns found from the
result. \figurename\ref{fig:unique-patterns}(a) shows that there were
not common ssh incoming activities.
\figurename\ref{fig:unique-patterns}(b) was observed when one of the
nodes in the target node group rebooted.

There are many approaches to detect anomalies or cluster messages
based-on templates
(\cite{xu2009-mining-console-logs,kimura2015-failure-detection,kimura2014-spatio-temporal-factorization-of-log,qiu2010-mining-events-from-router-syslog}).
Our template mining technology can be used with these
detection/clustering methods.

\section{Remaining Issues}

In this section we discuss issues of our proposed method.  Some of the
issues discussed here are not specific to our proposal only, but are
applied to online template mining methods in general.

The proposed algorithm doesn't take into account frequency of
appearance of words.  This causes state messages invisible from
output. For example, \texttt{interface eth0 up} and \texttt{interface
  eth1 down} may generate a template such as \texttt{interface * *}.
However, we may want two different templates \texttt{interface * up}
and \texttt{interface * down} in some cases.  Because we are focusing
on online realtime template generation, it is difficult to predict a
specific word is going to be a stable word like \texttt{up} and
\texttt{down} or not.

How to determine the threshold value is an important factor in the
method.  If the value is too loose, the algorithm will generate more
specific templates that will separate messages of same meaning to
different groups.  In this paper, we used 0.9 as $T_{c}$ and 3 as
$T_{p}$ to cluster three different message sources
(Table~\ref{tab:syslog-datasets}) achieved from different
administrative groups.  We could achieve the similar number of
templates as SHISO could infer with the threshold values when applied to
the standard Linux server syslog messages, however the proper values
may be different when applied to other kinds of dataset.

% XXX Overfitting issue

\section{Conclusion}

In this paper, we have proposed a new clustering method for inferring
system log message templates using the length of each words of
messages.  Many existing template mining approaches try to
characterize words of messages by their character types, ratio of
character types.  Our proposal comes from the question that do we
really need to investigate the word property so close.  We have
focused on the length of each word and found each message template
has a unique sequence of word length that can be used to cluster
messages.

The proposed method is designed for use of online (one-pass) template
mining.  The two-pass methods usually generate better templates by
surveying frequency of words to detect if a specific word is a fixed
word or a variable word, however they require time before clustering
and has difficulty to adapt dynamic changes of message trends.
Recently many systems are implemented with dynamic components such as
open source software.  Such components are updated continuously, and
even replaced with a different components providing the same
functionality.  In such a situation, adapting upgraded components
and/or new components are important.  Our proposed mechanism could
produce similar number of templates as past works with less complicity
of mining processing.

%\bibliographystyle{abbrv}
%\bibliographystyle{unsrt}
%\bibliography{bibliography.bib}

\end{document}